\documentclass{eptcs}
\usepackage{breakurl}
\providecommand{\event}{SOS 2012}

\newcommand\codett[1]{{\normalfont{\texttt{#1}}}}

\newcommand\handleexn[3]{\codett{try}\; #1 \; \codett{handle}\; #2 \; \codett{by}\; #3 } 
\newcommand\sep{\mathbin *}
\newcommand\sigblocks[2]{\normalfont\codett{block}\,#1\,\codett{in}\,#2} 
\newcommand\blocksonce[2]{\normalfont\codett{block/1}\,#1\,\codett{in}\,#2} 
\newcommand\sigbind[3]{\normalfont\codett{bind}\,#1\,\codett{to}\,#3\,\codett{in}\,#2} 
\newcommand\sigonce[3]{\normalfont\codett{bind/1}\,#1\,\codett{to}\,#3\,\codett{in}\,#2} 

\newcommand\seqcomp{\mathbin{\mathbf ;}}

\newcommand\infern[3]{\begin{prooftree} #1 \justifies #2 \using (#3) \end{prooftree}}
\newcommand\infer[2]{\begin{prooftree} #1 \justifies #2  \end{prooftree}}

\newcommand\mytheoremdefs{
\usepackage{theorem}
\theorembodyfont{\rm}
\newtheorem{definition}{Definition}[section]

}

\newcommand\step[1]{\stackrel{#1}\leadsto}

\newcommand\assign[2]{#1\,\codett{:=}\,#2}
\newcommand\throwexn[1]{\codett{throw}\,#1}

\newcommand\dom[1]{\mathrm{dom}(#1)}
\newcommand\update[3]{#1\,[\, #2 \mapsto #3 \,]}
\newcommand\hoaretriple[3]{ \{ #1 \} \, #2 \, \{ #3 \} } 
   
\newcommand\bs[5]{#1;#2 \vdash #3, #4 \mathrel\Downarrow #5 } 
\newcommand\bse[6]{#1;#2 \vdash #3, #4 \mathrel\Uparrow #5, #6 }
\newcommand\ebse[4]{#1, #2 \mathrel\Uparrow #3, #4 }
\newcommand\whiledo[2]{\codett{while}\,(#1)\,\codett{do}\,#2}

\newcommand\cekhb[5]{\blue\langle \mkern2mu#1 \mkern6mu \mathrel{\blue\mid} #2  \mathrel{\blue\mid} #3 \mathrel{\blue\mid} #4 \mathrel{\blue\mid} #5 \mkern2mu \blue\rangle}

\newcommand\blue[1]{\textcolor{blue}{#1}}

\newcommand\finalcont{\blacksquare}
\newcommand\return{\blue{\textbf{\texttt{return}}}}
\newcommand\sqplus[1]{\red{+#1}}
\newcommand\sqminus[1]{\red{-#1}}
\newcommand\stag[2]{(#1,#2)}

\newcommand\ttag[3]{(#1,#2,#3)}

\newcommand\popupd[1]{\blue{\normalfont\texttt{pop-upd(}}#1\blue{\normalfont\texttt{)}}}

\newcommand\upd[1]{\blue{\normalfont\texttt{update(}}#1\blue{\normalfont\texttt{)}}}

\newcommand\machsep{,\mkern3mu}

\mytheoremdefs

\newcommand\subsect[1]{\noindent\textbf{#1}}
\renewcommand\infern[3]{\infer{#1}{#2}} 

\newdimen\proofrulebreadth \proofrulebreadth=.05em
\newdimen\proofdotseparation \proofdotseparation=1.25ex
\newdimen\proofrulebaseline \proofrulebaseline=2ex
\newcount\proofdotnumber \proofdotnumber=3
\let\then\relax
\def\hfi{\hskip0pt plus.0001fil}
\mathchardef\squigto="3A3B
\newif\ifinsideprooftree\insideprooftreefalse
\newif\ifonleftofproofrule\onleftofproofrulefalse
\newif\ifproofdots\proofdotsfalse
\newif\ifdoubleproof\doubleprooffalse
\let\wereinproofbit\relax
\newdimen\shortenproofleft
\newdimen\shortenproofright
\newdimen\proofbelowshift
\newbox\proofabove
\newbox\proofbelow
\newbox\proofrulename
\def\shiftproofbelow{\let\next\relax\afterassignment\setshiftproofbelow\dimen0 }
\def\shiftproofbelowneg{\def\next{\multiply\dimen0 by-1 }%
\afterassignment\setshiftproofbelow\dimen0 }
\def\setshiftproofbelow{\next\proofbelowshift=\dimen0 }
\def\setproofrulebreadth{\proofrulebreadth}

\def\prooftree{
\ifnum  \lastpenalty=1
\then   \unpenalty
\else   \onleftofproofrulefalse
\fi
\ifonleftofproofrule
\else   \ifinsideprooftree
        \then   \hskip.5em plus1fil
        \fi
\fi
\bgroup
\setbox\proofbelow=\hbox{}\setbox\proofrulename=\hbox{}%
\let\justifies\proofover\let\leadsto\proofoverdots\let\Justifies\proofoverdbl
\let\using\proofusing\let\[\prooftree
\ifinsideprooftree\let\]\endprooftree\fi
\proofdotsfalse\doubleprooffalse
\let\thickness\setproofrulebreadth
\let\shiftright\shiftproofbelow \let\shift\shiftproofbelow
\let\shiftleft\shiftproofbelowneg
\let\ifwasinsideprooftree\ifinsideprooftree
\insideprooftreetrue
\setbox\proofabove=\hbox\bgroup$\displaystyle 
\let\wereinproofbit\prooftree
\shortenproofleft=0pt \shortenproofright=0pt \proofbelowshift=0pt
\onleftofproofruletrue\penalty1
}

\def\eproofbit{
\ifx    \wereinproofbit\prooftree
\then   \ifcase \lastpenalty
        \then   \shortenproofright=0pt  
        \or     \unpenalty\hfil         
        \or     \unpenalty\unskip       
        \else   \shortenproofright=0pt  
        \fi
\fi
\global\dimen0=\shortenproofleft
\global\dimen1=\shortenproofright
\global\dimen2=\proofrulebreadth
\global\dimen3=\proofbelowshift
\global\dimen4=\proofdotseparation
\global\count255=\proofdotnumber
$\egroup  
\shortenproofleft=\dimen0
\shortenproofright=\dimen1
\proofrulebreadth=\dimen2
\proofbelowshift=\dimen3
\proofdotseparation=\dimen4
\proofdotnumber=\count255
}

\def\proofover{
\eproofbit 
\setbox\proofbelow=\hbox\bgroup 
\let\wereinproofbit\proofover
$\displaystyle
}%
\def\proofoverdbl{
\eproofbit 
\doubleprooftrue
\setbox\proofbelow=\hbox\bgroup 
\let\wereinproofbit\proofoverdbl
$\displaystyle
}%
\def\proofoverdots{
\eproofbit 
\proofdotstrue
\setbox\proofbelow=\hbox\bgroup 
\let\wereinproofbit\proofoverdots
$\displaystyle
}%
\def\proofusing{
\eproofbit 
\setbox\proofrulename=\hbox\bgroup 
\let\wereinproofbit\proofusing
\kern0.3em$
}

\def\endprooftree{
\eproofbit 
  \dimen5 =0pt
\dimen0=\wd\proofabove \advance\dimen0-\shortenproofleft
\advance\dimen0-\shortenproofright
\dimen1=.5\dimen0 \advance\dimen1-.5\wd\proofbelow
\dimen4=\dimen1
\advance\dimen1\proofbelowshift \advance\dimen4-\proofbelowshift
\ifdim  \dimen1<0pt
\then   \advance\shortenproofleft\dimen1
        \advance\dimen0-\dimen1
        \dimen1=0pt
        \ifdim  \shortenproofleft<0pt
        \then   \setbox\proofabove=\hbox{%
                        \kern-\shortenproofleft\unhbox\proofabove}%
                \shortenproofleft=0pt
        \fi
\fi
\ifdim  \dimen4<0pt
\then   \advance\shortenproofright\dimen4
        \advance\dimen0-\dimen4
        \dimen4=0pt
\fi
\ifdim  \shortenproofright<\wd\proofrulename
\then   \shortenproofright=\wd\proofrulename
\fi
\dimen2=\shortenproofleft \advance\dimen2 by\dimen1
\dimen3=\shortenproofright\advance\dimen3 by\dimen4
\ifproofdots
\then
        \dimen6=\shortenproofleft \advance\dimen6 .5\dimen0
        \setbox1=\vbox to\proofdotseparation{\vss\hbox{$\cdot$}\vss}%
        \setbox0=\hbox{%
                \advance\dimen6-.5\wd1
                \kern\dimen6
                $\vcenter to\proofdotnumber\proofdotseparation
                        {\leaders\box1\vfill}$%
                \unhbox\proofrulename}%
\else   \dimen6=\fontdimen22\the\textfont2 
        \dimen7=\dimen6
        \advance\dimen6by.5\proofrulebreadth
        \advance\dimen7by-.5\proofrulebreadth
        \setbox0=\hbox{%
                \kern\shortenproofleft
                \ifdoubleproof
                \then   \hbox to\dimen0{%
                        $\mathsurround0pt\mathord=\mkern-6mu%
                        \cleaders\hbox{$\mkern-2mu=\mkern-2mu$}\hfill
                        \mkern-6mu\mathord=$}%
                \else   \vrule height\dimen6 depth-\dimen7 width\dimen0
                \fi
                \unhbox\proofrulename}%
        \ht0=\dimen6 \dp0=-\dimen7
\fi
\let\doll\relax
\ifwasinsideprooftree
\then   \let\VBOX\vbox
\else   \ifmmode\else$\let\doll=$\fi
        \let\VBOX\vcenter
\fi
\VBOX   {\baselineskip\proofrulebaseline \lineskip.2ex
        \expandafter\lineskiplimit\ifproofdots0ex\else-0.6ex\fi
        \hbox   spread\dimen5   {\hfi\unhbox\proofabove\hfi}%
        \hbox{\box0}%
        \hbox   {\kern\dimen2 \box\proofbelow}}\doll%
\global\dimen2=\dimen2
\global\dimen3=\dimen3
\egroup 
\ifonleftofproofrule
\then   \shortenproofleft=\dimen2
\fi
\shortenproofright=\dimen3
\onleftofproofrulefalse
\ifinsideprooftree
\then   \hskip.5em plus 1fil \penalty2
\fi
}

\usepackage[usenames,dvipsnames]{color}
\usepackage{hyperref}
\usepackage{amsmath}
\usepackage[english]{babel}
\usepackage[T1]{fontenc}

\usepackage{tikz}
\usetikzlibrary{decorations.pathreplacing}
\usepackage{amssymb}

\newcommand\m{-0.5ex}
\newcommand\n{-0.1ex}
\newcommand\red[1]{\textcolor{Red}{#1}}

\newcommand\unwind[3]{\texttt{unwind(}#1\machsep #2\machsep #3\texttt{)}}
\newcommand\eval[2]{\texttt{eval(}#1\machsep #2\texttt{)}}

\renewcommand\red[1]{#1}
\renewcommand\blue[1]{#1}

\title{Operational semantics for signal handling}

\author{Maxim Strygin
\institute{School of Computer Science\\
University of Birmingham\\ 
}
\email{M.Strygin@cs.bham.ac.uk}
\and
Hayo Thielecke
\institute{School of Computer Science\\
University of Birmingham\\ 
}
\email{\quad H.Thielecke@cs.bham.ac.uk }
}

\def\titlerunning{Operational semantics for signal handling}
\def\authorrunning{Strygin and Thielecke}
\begin{document}

\maketitle

\begin{abstract}
Signals are a lightweight form of interprocess communication in Unix. When a process receives a signal, the control flow is interrupted and a previously installed signal handler is run. Signal handling is reminiscent both of exception handling and concurrent interleaving of processes. In this paper, we investigate different approaches to formalizing signal handling in operational semantics, and compare them in a series of examples. We find the big-step style of operational semantics to be well suited to modelling signal handling. We integrate exception handling with our big-step semantics of signal handling, by adopting the exception convention as defined in the Definition of Standard ML. The semantics needs to capture the complex interactions between signal handling and exception handling.
\end{abstract}
\renewcommand\arraystretch{1.5}
 

\section{Introduction}

In operating systems, and specifically Unix and its descendants, signals provide a simple and efficient, if rather low-level, means of interprocess communication~\cite{Kerrisk:2010:LinuxProgInter,Robbins:2003:UNIX,Stevens:2005:APU,Loosemore:gnuClibRefMan,Bovet:2002:ULK}. 
Put simply, a process can cause a branch of control in another process, causing it to run a signal handler in response to external events. A well known example is the \texttt{kill} signal telling a process to shut down (perhaps after first deallocating system resources, such as releasing memory). 

Signals resemble exceptions in that control jumps to a handler that can be installed by the program.
Nonetheless, there are some significant differences. Whereas exceptions typically abort from the context, in which they were thrown rather than returning to it, signal handlers resume control after they have run. Whereas exceptions are triggered at specific points by the code itself, signals arrive nondeterministically. In the literature on control constructs and their semantics, signals have received less attention than exceptions, and far less than first-class continuations.

Exceptions have become amenable to semantic analysis by a focus on their key control features, while abstracting away from implementation details and restrictions (such as the entanglement of exceptions in C++ with the class hierarchy and memory management by destructors). For instance, the exceptions monad~\cite{moggi89computational} gives a highly idealized account of exceptions as functions $A \to (B+E)$ that may either return normally with a $B$ or raise an exception of type $E$.

The aim of the present paper is to address signal handling at a level of generality and abstraction comparable to that of other control constructs in the literature, idealizing where necessary and focusing on some key semantic features.
Our motivation for defining such a semantics, and exploring different styles of definition, is to develop of a Hoare logic for signals. While program logic is beyond the scope of the paper, it is a reason for our investigating the big-step style of operational semantics. In big-step, a command $c$ takes a pre-state $s_1$ to a post-state $s_2$ in a judgment of the form $s_1, c \Downarrow s_2$. This form of judgment is particularly convenient for proving the soundness of Hoare triples $\hoaretriple PcQ$, since the pre-condition $P$ refers to the pre-state $s_1$ and the postcondition $Q$ to the post-state $s_2$ in a big-step judgement.

\subsubsection*{Outline of the paper}

We begin by reviewing the constructs that we will need, and how to define operational semantics for them, in
Section~\ref{language}. We then combine these constructs and define the  semantics for the whole language
in Section~\ref{languageopsem}.
To validate our definition, we examine how signal and exception handling interact in a series of examples in
Section~\ref{secexamples}. As an alternative to big-step semantics, we define a small-step semantics as a stack machine in
Section~\ref{stackmachine}, and relate it to implementations. We  compare the stack machine to the big-step semantics in Section~\ref{machinerunsexamples}.
Section~\ref{conclusions} concludes.

\section{Language constructs} 
\label{language}

Before giving the formal definition of our operational semantics, we introduce the language constructs with their intended meaning, as well as design choices and simplifying assumptions. 
We start from a small imperative base language.
This language has a standard semantics in terms of how a command $c$ changes the state $s_{1}$ into a new state $s_{2}$. In a big-step operational semantics, the form of such judgements is
\[
s_{1}, c \Downarrow s_{2}
\]
When the command $c$ raises an exception $e$ after producing the new state $s_{2}$, we write 
\[
\ebse {s_{1}}ce{s_{2}}
\]

\subsect{Exceptions}

The semantics of exceptions is fairly well understood, and it is greatly simplified by the fact that exceptions are block structured. The more primitive non-local jumps in C (given via the library functions \texttt{setjmp()} and \texttt{longjmp()}) would be much harder to formalize.
Exception throwing and handling is easy to add to a big-step operational semantics. A classic example of such a semantics is the Definition of Standard ML~\cite{mlrevised}, whose style we will follow.

In addition to the rules for the operations themselves, we also need to specify how the propagation of exceptions interacts with the other constructs of the language: this propagation will be done with the \emph{exception convention} from the Definition of Standard ML.
If the $j$-th premise of a big-step rule raises an exception, and the premises to its left do not, then the conclusion of the rule raises the same exception, and with the same state.

More precisely, suppose there is a big-step rule of the form 
\[
\infer
{
\ldots c_{1} \Downarrow s_{1} \ldots c_{j} \Downarrow s_{j} \ldots c_{n} \Downarrow s_{n}
}
{\ldots c \Downarrow s
}
\]
Then we implicitly extend this case to propagating exception by adding a rule
\[
\infer
{
\ldots c_{1} \Downarrow s_{1} \ldots c_{j} \Uparrow e, s_{j} 
}
{\ldots c \Uparrow e,s_{j}
}
\]
To illustrate the exception convention, we consider how exceptions are propagated in a sequential composition
$c_{1};c_{2}$.
\[
\infer{
s_{1}, c_{1} \Uparrow e, s_{2}
}
{s_{1}, (c_{1};c_{2}) \Uparrow e, s_{2}}
\qquad
\infer{
s_{1}, c_{1} \Downarrow s_{2} \qquad s_{2},c_{2}\Uparrow e, s_{3} 
}
{s_{1}, (c_{1};c_{2}) \Uparrow e, s_{3}}
\]
 Intuitively, the first command $c_{1}$ may raise an exception, in which case the second command $c_{2}$ has not run at all. Alternatively, $c_{1}$ may terminate normally, and $c_{2}$ may raise an exception. In either case, the combined command raises the same exception.

\subsect{Signals}

The main construct we aim to address is signal handling. Signal handling is a form of interprocess communication, so that for full generality we would have to address the concurrent interaction between a signal sending and a signal handling process. To keep the semantics as simple as possible, we address only the \emph{handling} part of the signal mechanism, while the truly concurrent interaction between sender and receiver is left for future work. Rather than modelling the signal sender explicitly, only the point of view of the process receiving the signals will be assumed, so that signals arrive nondeterministically, causing handlers to run unpredictably. In the authors' view, this focus on signal \emph{handling} still presents sufficient programming and semantics challenges. First, the nondeterministic interference by signal handlers leads to the need to preserve resource invariants, much as interference between concurrent processes. Moreover, the assumptions a programmer can make about the delivery of signals are very weak, even if there is a specification
of the sender's behaviour (which there usually is not).
 In the worst case, the signal sender may even be malicious, sending signals with the sole intent of causing damage via the actions of the signal handlers. In that sense, a nondeterministic sender is a worst-case but realistic assumption that the signal receiver has to be able to cope with.

As a language construct, signal handlers resemble both concurrency and exception handling.
Our most significant idealization of signal handlers is directly inspired by exceptions in contrast to the unstructured \texttt{longjmp} that exceptions were designed to replace.
We define an idealized block-structured form of signal handling in which a signal handler is installed at the beginning of the block and uninstalled at the end. It relates to \texttt{sigaction} the way exceptions related  to \texttt{setjmp} and atomic \texttt{synchronized} blocks related to locking and unlocking. 

For the operational semantics, we define a big-step semantics. This style of semantics appears particularly apt for the signals and exceptions kind of constructs. Essentially, the meaning of a block becomes a subtree of a larger derivation tree, which is convenient for keeping track of pre- and post-states. In the same way, the derivation tree of one-sided signal handler could be easily injected into a larger tree.

One may think of addressing one-sided interleaving with the same approach as complete interleaving. This is true to some extent, but there are important differences between them. The interaction between fully concurrent processes is symmetric, but there is no such symmetry between the signal body and the handler.
Only the signal handler may interrupt the body, but not vice versa. This allows using a simpler approach for addressing signal handling. On the other hand, the general approach used for the fully concurrent interleaving might not be  suitable, as the interaction is non-symmetric.

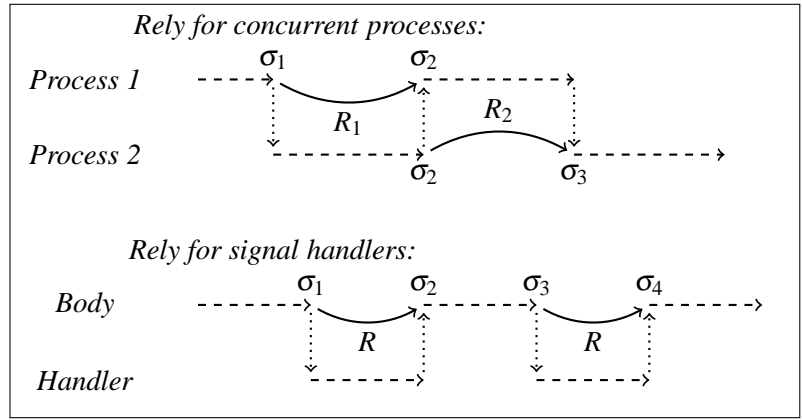
\begin{figure*}
\center
\begin{tikzpicture}

\coordinate (p1) at (0,0);
\coordinate (p2) at (1,0);
\coordinate (p3) at (3,0);
\coordinate (p4) at (5,0);

\coordinate (p5) at (0,-1);
\coordinate (p6) at (1,-1);
\coordinate (p7) at (3,-1);
\coordinate (p8) at (5,-1);
\coordinate (p9) at (7,-1);

\coordinate (p0) at (3,0.7);

\draw[->,dashed,thick] (p1) -- (p2);
\draw[->,dashed,thick] (p3) -- (p4);
\draw[->,dashed,thick] (p6) -- (p7);
\draw[->,dashed,thick] (p8) -- (p9);

\node[xshift=-15mm,yshift=0mm,font=\itshape] at (p0) {Rely for concurrent processes:};
\node[xshift=-15mm,yshift=0mm,font=\itshape] at (p1) {Process 1};
\node[xshift=-15mm,yshift=0mm,font=\itshape] at (p5) {Process 2};

\draw [->,dotted,thick,shorten <=3pt,shorten >=3pt] (p2) to (p6);
\draw [->,dotted,thick,shorten <=3pt,shorten >=3pt] (p7) to (p3);
\draw [->,dotted,thick,shorten <=3pt,shorten >=3pt] (p4) to (p8);

\draw [->,thick,shorten <=3pt,shorten >=3pt] (p2) to [bend left=-30] node [auto,swap] {$R_{1}$} (p3);
\draw [->,thick,shorten <=3pt,shorten >=3pt] (p7) to [bend left=+30] node [auto] {$R_{2}$} (p8);

\node[above] at (p2) {$\sigma_{1}$};
\node[above] at (p3) {$\sigma_{2}$};
\node[below] at (p7) {$\sigma_{2}$};
\node[below] at (p8) {$\sigma_{3}$};

\coordinate (b1) at (0,-3);
\coordinate (b2) at (1.5,-3);
\coordinate (b3) at (3,-3);
\coordinate (b4) at (4.5,-3);
\coordinate (b5) at (6,-3);
\coordinate (b6) at (7.5,-3);

\coordinate (h1) at (0,-4);
\coordinate (h2) at (1.5,-4);
\coordinate (h3) at (3,-4);
\coordinate (h4) at (4.5,-4);
\coordinate (h5) at (6,-4);

\coordinate (bh0) at (2.5,-2.3);

\draw[->,dashed,thick] (b1) -- (b2);
\draw[->,dashed,thick] (b3) -- (b4);
\draw[->,dashed,thick] (b5) -- (b6);
\draw[->,dashed,thick] (h2) -- (h3);
\draw[->,dashed,thick] (h4) -- (h5);

\node[xshift=-15mm,yshift=0mm,font=\itshape] at (bh0) {Rely for signal handlers:};
\node[xshift=-15mm,yshift=0mm,font=\itshape] at (b1) {Body};
\node[xshift=-15mm,yshift=0mm,font=\itshape] at (h1) {Handler};

\draw [->,dotted,thick,shorten <=3pt,shorten >=3pt] (b2) to (h2);
\draw [->,dotted,thick,shorten <=3pt,shorten >=3pt] (h3) to (b3);
\draw [->,dotted,thick,shorten <=3pt,shorten >=3pt] (b4) to (h4);
\draw [->,dotted,thick,shorten <=3pt,shorten >=3pt] (h5) to (b5);

\draw [->,thick,shorten <=3pt,shorten >=3pt] (b2) to [bend left=-30] node [auto,swap] {$R$} (b3);
\draw [->,thick,shorten <=3pt,shorten >=3pt] (b4) to [bend left=-30] node [auto,swap] {$R$} (b5);

\node[above] at (b2) {$\sigma_{1}$};
\node[above] at (b3) {$\sigma_{2}$};
\node[above] at (b4) {$\sigma_{3}$};
\node[above] at (b5) {$\sigma_{4}$};

\draw (-2.5,1) rectangle (8,-4.5);

\end{tikzpicture}
\caption{Processes vs signal handlers}%
\label{processvshandlers}%
\end{figure*}

Figure~\ref{processvshandlers} depicts the symmetric interleaving of concurrent processes compared to  
the one-sided interleaving of a process by its signal handlers. Dashed horizontal lines represent control flow; dotted vertical lines represent switches in the control flow due to interleaving.
In both cases, the state $\sigma_{i}$ that a process sees at some point could has been changed to some state $\sigma_{i+1}$ by interleaved actions. These state changes need to be limited in some way, as otherwise no assumptions could be made by the process about the state, including resource invariants.

\subsect{One-shot and persistent signals}

Signal handlers can have two different control flow semantics, which we call \emph{persistent} and \emph{one-shot}. 
A persistent signal handler can be run any number of times as long as it is installed. By contrast, a one-shot signal handler can be run at most once, as it becomes automatically uninstalled after being run the first time. In Unix, the system call for installing handlers takes a parameter that determines which of these behaviours is chosen.

\subsect{Operational semantics}

In the operational semantics, the evaluation of a command $c$ starting from a state $s_{1}$ will now take place relative to a signal binding. Moreover, the signal binding is subdivided into two parts: persistent signals $S$, and one-shot signals $O$. Persistent handlers may run any number of times during the evaluation of the command $c$, whereas one-shot handlers may run at most once.
The form of a big-step judgement with  signal bindings is:
\[
\bs SO{s_{1}}c{s_{2}}
\]
Note that the signal binding behaves like an environment (for variables bound via \texttt{let}) rather than a mutable state (for variables updated via \texttt{:=}). The judgement produces an updated state $s_{2}$, but it does not update $S$ or $O$.

Analogous to binding an exception handler, we have two binding constructs for signals: one for persistent and one for one-shot handlers, where $z$ is a signal name, $c_{b}$ is a command, and $c_{h}$ is a handler command.
\[
{\sigbind z{c_{b}}{c_{h}}}
\quad\mbox{ and }\quad
{\sigonce z{c_{b}}{c_{h}}}
\]
To support signal disabling in a scope, we introduce two blocking constructs for signals:
\[
\sigblocks{z}{c_{b}}
\quad\mbox{ and }\quad
\blocksonce{z}{c_{b}}
\]
Note that there is no need for an analogue of $\throwexn e$ (a command that throws an exception $e$), as we assume that signals arrive nondeterministically from other, unspecified processes.
The idea of using two contexts with a binder for each is loosely inspired by Barber and Plotkin's Dual Intuitionistic Linear Logic (DILL)~\cite{dill}.

\section{Operational semantics for block-structured signals and exceptions} 
\label{languageopsem}

\begin{definition}
\label{languagedef}
The syntax of the language with signal and exception handling is given in Figure~\ref{syntaxofthelanguage}.
\end{definition}

\begin{figure}
\[
\begin{array}{rcll} 
c &::=& \whiledo Ec & \mbox{(while construct)} 
\\
& \mid & x := E & \mbox{(Assignment)} 
\\
& \mid &  c_{1}; c_{2} & \mbox{(Sequential composition)} 
\\
& \mid & {\throwexn e} & \mbox{(Exception throwing)}
\\
& \mid & \handleexn{c_{1}}{e}{c_{2}} & \mbox{(Exception handling)}
\\
& \mid & \sigbind{z}{c_{2}}{c_{1}} & \mbox{(Binding persistent signal handler)}
\\
& \mid & \sigonce{z}{c_{2}}{c_{1}} & \mbox{(Binding one-shot signal handler)}
\\
& \mid & \sigblocks{z}{c} & \mbox{(Blocking persistent signal)}
\\
& \mid & \blocksonce{z}{c} & \mbox{(Blocking one-shot signal)}
\\
E &::=& x \mid E+E \mid \ldots & \mbox{(Expressions)}
\end{array}
\]
\caption{The syntax of the language}
\label{syntaxofthelanguage}
\end{figure}

\begin{figure} 

\[
\begin{array}{c@{\hspace{2em}}c}
\infern{
\bs{\update{S}{z}{c_{1}}}{O}{s_{1}}{c_{2}}{s_{2}}
}
{
\bs{S}{O}{s_{1}}{\sigbind{z}{c_{2}}{c_{1}}}{s_{2}}
}{\textsc{PerSigBind}}
&
\infern{
\bs{S}{\update{O}{z}{c_{1}}}{s_{1}}{c_{2}}{s_{2}}
}
{
\bs{S}{O}{s_{1}}{\sigonce{z}{c_{2}}{c_{1}}}{s_{2}}
}{\textsc{OneSigBind}}
\\[2em]
\infern{
\bs{S-z}{O}{s_{1}}{c}{s_{2}}
}
{
\bs{S}{O}{s_{1}}{\sigblocks{z}{c}}{s_{2}}
}{\textsc{PerSigBlock}}
&
\infern{
\bs{S}{O-z}{s_{1}}{c}{s_{2}}
}
{
\bs{S}{O}{s_{1}}{\blocksonce{z}{c}}{s_{2}}
}{\textsc{OneSigBlock}}
\end{array}
\]
\[
\infern{}{
\bse SO{s}{\throwexn e}e{s}
}{\textsc{Throw}}
\]
\\[\m]
\[
\begin{array}{c@{\hspace{2em}}c}
\infern{
\bse S{O_{1}}{s_{1}}{c_{1}}{e}{s_{2}}
\qquad
\bs S{O_{2}}{s_{2}}{c_{2}}{s_{3}}
}
{
\bs S{O_{1} \sep O_{2}}{s_{1}}{\handleexn{c_{1}}{e}{c_{2}}}{s_{3}}
}{\textsc{Handl}}
&
\infern{
\bs SO{s_{1}}{c_{1}}{s_{2}}
}
{
\bs SO{s_{1}}{\handleexn{c_{1}}{e}{c_{2}}}{s_{2}}
}{\textsc{HANDL2}}
\\[2em]
\infern{
\bse S{O_{1}}{s_{1}}{c_{1}}{e}{s_{2}}
\qquad
\bse S{O_{2}}{s_{2}}{c_{2}}{e_{2}}{s_{3}}
}
{
\bse S{O_{1} \sep O_{2}}{s_{1}}{\handleexn{c_{1}}{e}{c_{2}}}{e_{2}}{s_{3}}
}{\textsc{HANDL3}}
&
\infern{
\bse SO{s_{1}}{c_{1}}{e_{2}}{s_{2}}
\qquad e_{2} \neq e
}
{
\bse SO{s_{1}}{\handleexn{c_{1}}{e}{c_{2}}}{e_{2}}{s_{2}}
}{\textsc{HANDL4}}
\end{array}
\]
\\[\m]
\[
\begin{array}{c@{\hspace{2em}}c}
\infern{s\vdash E\downarrow v}
{
\bs SO{s}{\assign xE}{\update sxv}
}{\textsc{Atomic}}
&
\infern{
\bs S{O_{1}}{s_{1}}{c_{1}}{s_{2}}
\qquad
\bs S{O_{2}}{s_{2}}{c_{2}}{s_{3}}
}{
\bs S{O_{1} \sep O_{2}}{s_{1}}{(c_{1};c_{2})}{s_{3}}
}{\textsc{SeqComp}}
\end{array}
\]
\\[\m]
\[
\begin{array}{c@{\hspace{2em}}c}
\infern{
\bs {S}{O}{s_{1}}{c_{1}}{s_{2}}
\quad
S(z)=c_{2}
\quad
\bs{\emptyset}{\emptyset}{s_{2}}{c_{2}}{s_{3}}
}
{\bs SO{s_{1}}{c_{1}}{s_{3}}}{\textsc{PerShotHandl}}
&
\infern{
\bs S{O-z}{s_{1}}{c_{1}}{s_{2}}
\quad
O(z)=c_{2}
\quad
\bs{\emptyset}{\emptyset}{s_{2}}{c_{2}}{s_{3}}
}
{\bs SO{s_{1}}{c_{1}}{s_{3}}}{\textsc{OneShotHandl}}
\\[2em]
\infern{
S(z)=c_{2}
\quad
\bs{\emptyset}{\emptyset}{s_{1}}{c_{2}}{s_{2}}
\quad
\bs {S}{O}{s_{2}}{c_{1}}{s_{3}}
}
{\bs SO{s_{1}}{c_{1}}{s_{3}}}{\textsc{PerShotHandl2}}
&
\infern{
O(z)=c_{2}
\quad
\bs{\emptyset}{\emptyset}{s_{1}}{c_{2}}{s_{2}}
\quad
\bs S{O-z}{s_{2}}{c_{1}}{s_{3}}
}
{\bs SO{s_{1}}{c_{1}}{s_{3}}}{\textsc{OneShotHandl2}}
\end{array}
\]
\caption{Big-step semantics rules for exceptions and signal handling}
\label{figbsrulesfexnsign}
\end{figure}

We let $s$ range over states, $c$ over commands, $x$ over variables, $v$ over values, $e$ over exception names, $z$ over signal names, and $E$ over expressions, using subscripts where needed, e.g., $s_{1}$, $e_{2}$, $E_{3}$, $c_{4}$ or $c_{h}$.
$s$ is a function from variables to values, such that $s(x)$ returns a value $v$.

Some auxiliary definitions will be required for the operational semantics.
For a partial function $f$, we write $\update fxv$ for the function that maps $x$ to $v$ and coincides with
$f$ on all other arguments. In particular, we use this notation for updating states or signal bindings. We write $\dom f$ for the domain of definition of a partial function. For $x\in \dom f$, we write $f-x$ for the restriction of $f$ to $(\dom f \setminus \{x\})$.
A signal binding is a finite partial function from signal names $z$ to commands $c$.
We will need a partial operation on signal bindings. In fact, this definition is the same as
the separating conjunction from separation logic~\cite{reynoldslicssep}.
\begin{definition}
\label{sepoperation}
Given two signal bindings $O_{1}$ and $O_{2}$, we define a partial operation $\sep$ as follows:
\begin{itemize}
\item
 If $\dom{O_{1}} \cap \dom{O_{2}} = \emptyset$, we write $O_{1} \sep O_{2}$ for $O_{1} \cup O_{2}$. 
 \item
 If $\dom{O_{1}} \cap \dom{O_{2}} \neq \emptyset$, then  $O_{1} \sep O_{2}$ is undefined.
 \end{itemize}
\end{definition}
 It is this splitting of a signal binding, analogous to the heap-splitting of separation logic, that gives one-shot behaviour to signals. Specifically, in a sequential composition $(c_{1}; c_{2})$, the one-shot signals are split non-deterministically between the commands $c_{1}$ and $c_{2}$. 
Moreover, every time a one-shot signal arrives and is handled, it is removed from the one-shot binding O. Thus, a one-shot signal may never be handled twice.
 
\begin{definition}
\label{opsem}
Given two signal bindings $S$ and $O$, the form of a big-step judgement is either 
\[
\bs SO{s_{1}}c{s_{2}}
\]
for normal termination, or
\[
\bse SO{s_{1}}ce{s_{2}}
\]
for exception throwing. The rules are given in
Figure~\ref{figbsrulesfexnsign}. 
The exception convention is assumed implicitly.
\end{definition}

\section{Examples}
\label{secexamples}
We examine how signal and exception handling interact in a series of examples, and discuss the question of priority between them.

\begin{figure}
\[
\infer{
  \infer{
  	\infer{
		\update{O_{1}}{z}{c_{h}}(z) = c_{h}
		\qquad
		\bs{\emptyset}{\emptyset}{s_{1}}{c_{h}}{s_{2}}
		\qquad
		\bs{S}{O_{1}-z}{s_{2}}{c_{1}}{s_{3}} 
	}{
  		\bs {S}{\update{O_{1}}{z}{c_{h}}}{s_{1}}{c_{1}}{s_{3}}
   	}
   \qquad 
   \bs {S}{O_{2}}{s_{3}}{c_{2}}{s_{4}}
  }{
   \bs {S}{\update{(O_{1} \sep O_{2})}{z}{c_{h}}}{s_{1}}{(c_{1} \seqcomp c_{2})}{s_{4}}
  } 
}{
\bs S{O_{1} \sep O_{2}}{s_{1}}{\sigonce{z}{(c_{1} \seqcomp c_{2})}{c_{h}}}{s_{4}}
}
\]
\caption{Splitting of the $O$ binding in seq. composed commands}
\label{oneshotsigomegasplit}
\end{figure}

\begin{figure}
\[
\infer{
 \infer{
	\infer{
	 \update{S}{z}{c_{h}}(z)=c_{h}
	 \qquad
	 \bs{\emptyset}{\emptyset}{s_{1}}{c_{h}}{s_{2}}
	 \qquad
	 \bs {\update{S}{z}{c_{h}}}{O}{s_{2}}{c_{1}}{s_{3}}
	}{
	 \bs{\update{S}{z}{c_{h}}}{O}{s_{1}}{c_{1}}{s_{3}}
	}
	\qquad
	\mathcal{D}
 }{
  \bs{\update{S}{z}{c_{h}}}{O}{s_{1}}{(c_{1} \seqcomp c_{2})}{s_{6}}
 }
}{
 \bs{S}{O}{s_{1}}{\sigbind{z}{(c_{1} \seqcomp c_{2})}{c_{h}}}{s_{6}}
}
\]
\[
\begin{array}{cc}
\mathcal{D} = & 
\\[\m]
&\infer{
	 \update{S}{z}{c_{h}}(z)=c_{h}
	 \qquad
	 \bs{\emptyset}{\emptyset}{s_{3}}{c_{h}}{s_{4}}
	 \qquad
	 \mathcal{F}
	}{
	 \bs{\update{S}{z}{c_{h}}}{O}{s_{3}}{c_{2}}{s_{6}}
	}
\end{array}
\]
\[
\begin{array}{cc}
\mathcal{F} = & 
\\[\m]
&
\infer{
 \update{S}{z}{c_{h}}(z)=c_{h}
 \qquad
 \bs{\emptyset}{\emptyset}{s_{4}}{c_{h}}{s_{5}}
 \qquad
 \bs {\update{S}{z}{c_{h}}}{O}{s_{5}}{c_{2}}{s_{6}}
}{
 \bs {\update{S}{z}{c_{h}}}{O}{s_{4}}{c_{2}}{s_{6}}
}
\end{array}
\]
\caption{Multiple persistent signal handling in seq. composed commands}
\label{seqcompsighandlcomp}
\end{figure}

\subsect{Examples for signals}

The aim of the Figure~\ref{oneshotsigomegasplit} and Figure~\ref{seqcompsighandlcomp} is to show how one-shot and persistent signal bindings are "shared" between sequentially composed commands, and highlight the core difference between them (splitting versus copying).

In Figure~\ref{oneshotsigomegasplit}, the one-shot signal binding $O = O_{1} \sep O_{2}$ (Definition~\ref{sepoperation}) is split non-deterministically between commands $c_{1}$ and $c_{2}$.
When the new signal $z$ is registered, it becomes an element of the domain $\update{(O_{1} \sep O_{2})}{z}{c_{h}}$. However, is $z \in \dom{\update{O_{1}}{z}{c_{h}}}$ or $z \in \dom{\update{O_{2}}{z}{c_{h}}}$ will be determined during the run time only.
In this particular example, the signal $z$ arrives in "scope" of the command $c_{1}$ ($z \in \dom{\update{O_{1}}{z}{c_{h}}}$) and the bound handler runs. According to the one-shot signal binding nature, the binding for $z$ is removed from  $\update{O_{1}}{z}{c_{h}}$ and consequently from  $\update{(O_{1} \sep O_{2})}{z}{c_{h}}$ as $\update{O_{1}}{z}{c_{h}} \subseteq \update{(O_{1} \sep O_{2})}{z}{c_{h}}$.
Therefore, $z \notin \dom{O_{2}}$ and if  the signal $z$ arrives during the execution of the command $c_{2}$, it will be ignored.

In Figure~\ref{seqcompsighandlcomp}, we focus on a persistent signal binding. The key difference with the one-shot binding is that the binding is just copied to the every command without splitting or modification. 
Thus, the same signal handler may run any number of times during the execution of the commands $c_{1}$ and $c_{2}$. This behaviour is possible because triggering a persistent signal handler does not remove the corresponding binding.

\subsect{Examples for signals and exceptions}

Suppose that a signal handler relies on some resource (valid pointer, open socket, active connection, etc.) available in the particular scope. However, as a side effect of the handler execution, the resource becomes unavailable (freed pointer, closed socket, inactive connection). In this situation, multiple handler executions may lead to the program fail and abrupt termination.

Obviously, one-shot signal handlers are perfectly fit for purpose. In Figure~\ref{oneshotsignhandlbasic}, the one-shot signal handler $c_{h}$ runs before the command $c$. Thus, when control flow returns to $c$, the one-shot signal binding no longer contains a binding for the handler $c_{h}$. 
In Figure~\ref{oneshotsignhandlbasic2}, the one-shot signal handler $c_{h}$ runs after the command $c$, and  at that point the signal binding no longer contains a binding for $c_{h}$. Note that the signal handlers (persistent and one-shot) that are still bound might be triggered if the corresponding signal arrives after the $c_{h}$.

\begin{figure}
\[
  \infer{
  	\infer{
		\update{O}{z}{c_{h}}(z) = c_{h}
		\qquad
		\bs{\emptyset}{\emptyset}{s_{1}}{c_{h}}{s_{2}}
		\qquad
		\bs{S}{O-z}{s_{2}}{c}{s_{3}} 
	}{
   		\bs {S}{\update{O}{z}{c_{h}}}{s_{1}}{c}{s_{3}}
   	}
  }{
   \bs S{O}{s_{1}}{\sigonce{z}{c}{c_{h}}}{s_{3}}
  } 
\]
\caption{One-shot signal handling before the command}
\label{oneshotsignhandlbasic}
\end{figure}

\begin{figure}
\[
  \infer{
  	\infer{
		\bs{S}{O-z}{s_{1}}{c}{s_{2}}
		\qquad
		\update{O}{z}{c_{h}}(z) = c_{h}
		\qquad
		\bs{\emptyset}{\emptyset}{s_{2}}{c_{h}}{s_{3}}		 
	}{
   		\bs {S}{\update{O}{z}{c_{h}}}{s_{1}}{c}{s_{3}}
   	}
  }{
   \bs S{O}{s_{1}}{\sigonce{z}{c}{c_{h}}}{s_{3}}
  } 
\]
\caption{One-shot signal handling after the command}
\label{oneshotsignhandlbasic2}
\end{figure}

On the other hand, a persistent handler combined with an exception imitates one-shot signal handlers to some extent. The key trick is in adding of a "$\throwexn{}$" command to the end of the persistent handler. Because of a thrown exception, control leaves the signal block, so the persistent signal handler will not run again.

In Figure~\ref{perssignhandlbefcom}, the persistent signal handler runs and throws an exception. As exception propagation takes place,  the command $c$ does not run.
In Figure~\ref{perssignhandlaftcom}, the command $c$ runs before the persistent signal handler has been triggered. Thus, the raising of the exception does not influence the command $c$ at that point.

Comparing the derivation trees in Figure~\ref{oneshotsignhandlbasic2} and Figure~\ref{perssignhandlaftcom}, we observe how similar they are. In both cases, the main command runs first and then the signal handler runs only once. The only difference is that  singular executions  of the handler has been achieved by two different approaches.

Comparing the derivation trees from Figure~\ref{oneshotsignhandlbasic} and Figure~\ref{perssignhandlbefcom}, we observe the next situation: in both cases the strict condition (singular execution) for the signal handlers is satisfied, but as a "side effect" of an exception propagation (Figure~\ref{perssignhandlbefcom}), the command $c$ is skipped.

\begin{figure}
\[
\infer{
 \infer{
  \infer{
   \update{S}{z}{(h \seqcomp\throwexn e)}(z) = (h \seqcomp\throwexn e) 
   \quad 
   \infer{
    \bs{\emptyset}{\emptyset}{s_{1}}{h}{s_{2}}
    \quad
    \infer{}{
    	\bse{\emptyset}{\emptyset}{s_{2}}{\throwexn e}{e}{s_{2}}
    }
   }{
    \bse{\emptyset}{\emptyset}{s_{1}}{(h \seqcomp\throwexn e)}{e}{s_{2}}
   } 
  }{
   \bse {\update{S}{z}{(h \seqcomp\throwexn e)}}{O}{s_{1}}{c}{e}{s_{2}}
  } 
 }{
  \bse SO{s_{1}}{(\sigbind{z}{c}{(h \seqcomp\throwexn e)})}{e}{s_{2}}
 }
 \quad \bs SO{s_{2}}{g}{s_{3}}
}{
 \bs SO{s_{1}}{\handleexn{(\sigbind{z}{c}{(h \seqcomp\throwexn e)})}{e}{g}}{s_{3}}
}
\]
\caption{Persistent handler with an exception triggered before the command}
\label{perssignhandlbefcom}
\end{figure}

\begin{figure}
\[
\infer{
 \infer{
  \infer{
   \bs {\update{S}{z}{(h \seqcomp\throwexn e)}}{O}{s_{1}}{c}{s_{2}}
   \quad
   \update{S}{z}{(h \seqcomp\throwexn e)}(z) = (h \seqcomp\throwexn e) 
   \quad 
   \mathcal{F} 
  }{
   \bse {\update{S}{z}{(h \seqcomp\throwexn e)}}{O}{s_{1}}{c}{e}{s_{3}}
  } 
 }{
  \bse SO{s_{1}}{(\sigbind{z}{c}{(h \seqcomp\throwexn e)})}{e}{s_{3}}
 }
 \quad \bs SO{s_{3}}{g}{s_{4}}
}{
 \bs SO{s_{1}}{\handleexn{(\sigbind{z}{c}{(h \seqcomp\throwexn e)})}{e}{g}}{s_{4}}
}
\]
\[
\begin{array}{cc}
\mathcal{F} = & 
\\[\m]
&\infer{
    \bs{\emptyset}{\emptyset}{s_{2}}{h}{s_{3}}
    \qquad
    \infer{}{
    	\bse{\emptyset}{\emptyset}{s_{3}}{\throwexn e}{e}{s_{3}}
    }
   }{
    \bse{\emptyset}{\emptyset}{s_{2}}{(h \seqcomp\throwexn e)}{e}{s_{3}}
   }
\end{array} 
\]
\caption{Persistent handler with an exception triggered after the command}
\label{perssignhandlaftcom}
\end{figure}

\begin{figure}
\[
\infer{
 \infer{
  \infer{
   \update{S}{z}{h}(z)=h
   \qquad
   \bs{\emptyset}{\emptyset}{s_{1}}{h}{s_{2}} 
   \qquad 
   \infer{}{
   	\bse {\update{S}{z}{h}}{O}{s_{2}}{\throwexn e}{e}{s_{2}}
   }
  }{\bse {\update{S}{z}{h}}{O}{s_{1}}{\throwexn e}{e}{s_{2}}}
 }{
  \bse SO{s_{1}}{(\sigbind{z}{\throwexn e}{h})}{e}{s_{2}}
 }
\quad
\bs SO{s_{2}}{g}{s_{3}}
}
{
\bs SO{s_{1}}{\handleexn{(\sigbind{z}{\throwexn e}{h})}{e}{g}}{s_{3}}
}
\]
\caption{Example of the derivation tree: a signal binding inside an exception block}
\label{example2sigexn}
\end{figure}

\subsect{Interaction between signal and exception handling}
\label{interaction}

There is potentially a pitfall in combining signals and jumps (such as exceptions), in that a jump could prevent a handler from being correctly uninstalled at the end of its scope.
 In fact the problem is quite general, and arises whenever resource management is combined with jumping.
In our language as defined in Definition~\ref{languagedef}, 
such a potential problem case is presented by the following code:
\[
\handleexn{(\sigbind{z}{\throwexn e}{h})}{e}{g}
\]
The intended meaning is that the signal $z$ is bound locally inside the body of an exception block. The signal handler may run immediately before the $\throwexn e$ command. 
However, once the exception has propagated to the exception handler, it has left the scope of the signal binding, so that the signal handler should not be able to run. To see that the big-step semantics (Figure~\ref{figbsrulesfexnsign}) correctly handles this case, consider the derivation tree in Figure~\ref{example2sigexn}.

In a big-step semantics, block structure is handled correctly "for free".
The extended signal binding $\update{S}{z}{h}$ is confined to the subtree of the body of the binding.
When the body is left, the evaluation is resumed with the old $S$, which is what is used in the evaluation of $g$. Even when control leaves the signal block abruptly via an exception, there is no danger that the signal handler escapes from its scope.
By contrast, in a small-step semantics (e.g.: abstract machine) the uninstalling of signal handlers needs to be performed explicitly.

\begin{figure}
\[
\infer{
 \infer{
   \infer{
   	\update{S}{z}{h}(z)=h
   	\qquad
   	\bs{\emptyset}{\emptyset}{s_{1}}{h}{s_{2}} 
   	\qquad 
   	\infer{}{
      \bse {\update{S}{z}{h}}{O}{s_{2}}{\throwexn e}{e}{s_{2}}
    }
   }{
   \bse {\update{S}{z}{h}}O{s_{1}}{\throwexn e}{e}{s_{2}}
   }
   \qquad
   \mathcal{F}
 }{
  \bs {\update{S}{z}{h}}O{s_{1}}{\handleexn{(\throwexn e)}{e}{g}}{s_{4}}
 }
}{
  \bs SO{s_{1}}{\sigbind{z}{(\handleexn{(\throwexn e)}{e}{g})}{h}}{s_{4}}
}
\]
\[
\begin{array}{cc}
\mathcal{F} = & 
\\[\m]
&\infer{
   	\update{S}{z}{h}(z)=h
   	\qquad
   	\bs{\emptyset}{\emptyset}{s_{2}}{h}{s_{3}} 
   	\qquad
	\bs {\update{S}{z}{h}}O{s_{3}}{g}{s_{4}}
   }{
   	\bs {\update{S}{z}{h}}O{s_{2}}{g}{s_{4}}
   }
\end{array} 
\]

\caption{Derivation tree for the combined signals and exceptions}
\label{exnpriorityoversign}
\end{figure}

\begin{figure}
\[
\infer{
 \infer{
  \infer{
    \infer{}{
  		\bse {\update{S}{z}{h}}{O}{s_{1}}{\throwexn e}{e}{s_{1}}
	}
	\qquad
	\red{\update{S}{z}{h}(z)=h}
  	\qquad
  	\red{\bs{\emptyset}{\emptyset}{s_{1}}{h}{s_{2}}}    
  }{\bse {\update{S}{z}{h}}{O}{s_{1}}{\throwexn e}{e}{s_{2}}}
 }{
  \bse SO{s_{1}}{(\sigbind{z}{\throwexn e}{h})}{e}{s_{2}}
 }
\qquad
\bs SO{s_{2}}{g}{s_{3}}
}
{
\bs SO{s_{1}}{\handleexn{(\sigbind{z}{\throwexn e}{h})}{e}{g}}{s_{3}}
}
\]
\caption{Signal handler runs after the \texttt{throw}}
\label{signpriorityoverexn}
\end{figure}

\subsect{The question of priority}

In our operational semantics, exception propagation has higher priority than exception handling.
Thus, a signal might be handled only before the exception has been \textit{thrown} and after it has been caught (Figure~\ref{exnpriorityoversign}). 
The command \texttt{throw} does not change the state itself, thus the state remains unchanged until the exception is caught, when there are different options: if no signal arrives then the exception handler runs, or else the signal handler runs first, and only then the exception handler proceeds.

However, one can design an implementation where signal handling has higher priority.
Thus, a signal handler should be processed even if exception propagation takes place (Figure~\ref{signpriorityoverexn}).
In a semantics with signal priority, the state is changed by the signal handler even during the exception propagation. One can make a few interesting observation about it. During exception propagation, control flow exits nested blocks, which in turn may have different signal bindings. Thus, depending in which block a signal arrives, the corresponding handler will interrupt the exception propagation. In addition, it might be the case that the signal is blocked in that scope, thus propagation would not be interrupted.

\section{Stack machine for signal handlers}
\label{stackmachine}

We define an abstract machine in order to highlight some of the issues that may arise in possible implementations of block-structured signals, such as managing the stack.
The implementation of signal handlers in our abstract machine was inspired by the real implementations of exceptions in contrast to the unstructured \texttt{longjmp} that exceptions were designed to replace. 

The defined block-structured form of signal handling requires a signal handler to be installed at the beginning of the block and uninstalled at the end.
Therefore, 
to keep track of signal handlers in a particular scope, we use a signal stack.
However, the addition of exceptions complicates the scoping of signal handlers. When control leaves a signal scope via a raised exception, the handler should be uninstalled.
Thus, to implement the desired interaction between signal and exception scope, we keep track of signal handlers and exception handlers on the same stack. When an exception is raised, the stack is popped until the nearest enclosing handler for the exception name is found. The same popping of the common handler stack also removes any intervening signal handlers.

A machine configuration is of the form \(\cekhb{c}{s}{\beta}{J}{K}\),  where
$c$ is the expression that the machine is currently trying to evaluate, 
$s$ is a state.
The bit vector component $\beta$ is used for keeping track of installed (not blocked) signals.
$J$ is a stack, which holds the signal and exception bindings.
$K$ is a continuation, which tells the machine what to do when it is finished with the current command $c$.
The initial continuation is a special instruction $\return$.
The special symbol $\finalcont$ is used to represent an empty stack in the components $J$ and $K$.
When we get $\cekhb{\return}{s}{\beta}{\finalcont}{\finalcont}$, program execution is finished. The full list of transition steps is given in Figure~\ref{blockstructsignlexnvbm82}.
To evaluate expression $E$ in a state $s$, we apply the function $eval$ (Defintion~\ref{evalfunctiondef}), which returns a value $v$.

$\beta^{0}$ stands for a null bit vector (which means blocking or ignoring of all signals).
The system instruction $\popupd{\beta'}$ removes the top element from a stack $J$ and updates $\beta$ to $\beta'$.
The system instruction $\upd{\beta'}$ updates $\beta$ to $\beta'$.
We define $J$ as a data structure that follows stack discipline except in the case of one-shot signal handling. The $J$ stack is manipulated by the system instructions that are pushed in and popped out from the continuation stack $K$.

$\beta$ is a function from signal names $z$ to \texttt{Booleans}. For each signal name $z$, $\beta(z)$ tells us whether the signal is currently enabled.
Then $\beta \sqplus z$ is a shorthand for $\beta[z \mapsto true]$ and $\beta \sqminus z$ stands for $\beta[z \mapsto false]$.

For a $\throwexn e_{1}$ command, where $e_{1} \in \dom{J}$, we apply the \texttt{unwind} function (Definition~\ref{unwindfunctiondef}), which returns a quadruple that is used to construct the next machine configuration.
If $e_{1} \notin \dom{J}$, then the machine gets stuck with an unhandled exception, in the sense that there is no transition for this configuration, so that
\[
	\cekhb{\throwexn e_{1}}{s}{\beta}{J}{K} \not\step{} 
\]

An exception binding tag has the form of $\stag{e}{h}$, where $e$ is an exception identifier, and $h$ is a handler. 
A persistent signal binding tag has the form of $\stag{z}{h}$, where $z$ is a signal name, and $h$ is a handler.
A one-shot signal binding tag has the form of $\ttag{z}{h}{u}$, where $z$ is a signal name, $h$ is a handler, and $u$ is a bit indicating that the handler has been used once ($u$=1) or not ($u$=0).
Handling of the one-shot signals requires update of the $J$ stack; to be more precise, the bit $u$ in $\ttag{z}{h}{u}$ is updated. 

\begin{figure}
\[
  \begin{array}{rcl}
  	\cekhb{c_{1};c_{2}}{s_{1}}{\beta_{1}}{J_{1}}{K_{1}}
	& \step{} & \cekhb{c_{1}}{s_{1}}{\beta_{1}}{J_{1}}{c_{2};K_{1}}
	\\[0.5ex]
	\cekhb{x := E}{s_{1}}{\beta_{1}}{J_{1}}{c';K_{1}} & \step{} &	\cekhb{c'}{\update {s_{1}}{x}{v}}{\beta_{1}}{J_{1}}{K_{1}}
	\\
	&& where \; \eval{E}{s_{1}}=v
	\\
	\cekhb{\sigbind{z}{c}{h}}{s}{\beta}{J}{K} & \step{} & \cekhb{c}{s}{\beta \sqplus z }{\stag{z}{h},J}{\popupd{\beta};K}
	\\[0.5ex]
	\cekhb{\sigonce{z}{c}{h}}{s}{\beta}{J}{K} & \step{} & \cekhb{c}{s}{\beta \sqplus z }{\ttag{z}{h}{0},J}{\popupd{\beta};K}
	\\[0.5ex]
	\cekhb{\popupd{\beta_{1}}}{s}{\beta_{2}}{\stag{z}{h},J}{c;K} & \step{} & \cekhb{c}{s}{\beta_{1}}{J}{K}
	\\[0.5ex]
	\cekhb{c}{s}{\beta}{J_{1},\stag{z}{h},J_{2}}{K} & \step{} & \cekhb{h}{s}{\beta^{0}}{J_{1},\stag{z}{h},J_{2}}{\upd{\beta};c;K}
	\\
	&& handling \; of \; the \; persistent \; signal
	\\
	\cekhb{c}{s}{\beta}{J_{1},\ttag{z}{h}{0},J_{2}}{K} & \step{} & \cekhb{h}{s}{\beta^{0}}{J_{1},\ttag{z}{h}{1},J_{2}}{\upd{\beta \sqminus z};c;K}
	\\
	&& handling \; of \; the \; one-shot \; signal
	\\
	\cekhb{\sigblocks{z}{c}}{s}{\beta_{1}}{J}{K} & \step{} & \cekhb{c}{s}{\beta_{1} \sqminus z}{J}{\upd{\beta_{1}};K}
	\\[0.5ex]
	\cekhb{\blocksonce{z}{c}}{s}{\beta_{1}}{J}{K} & \step{} & \cekhb{c}{s}{\beta_{1} \sqminus z}{J}{\upd{\beta_{1}};K}	
	\\[0.5ex]
	\cekhb{\upd{\beta_{1}}}{s}{\beta_{2}}{J}{c;K} & \step{} & \cekhb{c}{s}{\beta_{1}}{J}{K}
	\\[0.5ex]
\cekhb{\handleexn{c_{b}}{e}{h}}{s}{\beta}{J}{K} & \step{} & \cekhb{c_{b}}{s}{\beta}{\stag{e}{h},J}{\popupd{\beta};K}	
	\\[1ex]
	 \cekhb{\throwexn e_{1}}{s}{\beta}{J_{1},\stag{e_{1}}{h}, J_{2}}{K_{1}} & \step{} & \cekhb{h}{s}{\beta'}{J_{2}}{K_{2}}	 
  \end{array}
\]
\begin{flushright}
 where $\unwind{e_{1}}{(J_{1},\stag{e_{1}}{h}, J_{2})}{K_{1}}$ = $(h,\beta',J_{2},K_{2})$.
\end{flushright}

\caption{Transition steps}
\label{blockstructsignlexnvbm82}
\end{figure}

\begin{definition}[eval function]
\label{evalfunctiondef}
\[
\begin{array}{rcl} 
	\eval{x}{s} & = & s(x)
	\\
	\eval{E_{1}+E_{2}}{s} & = & \eval{E_{1}}{s} + \eval{E_{2}}{s}
\end{array}	
\]
\end{definition}

\begin{definition}[unwind function]
\label{unwindfunctiondef}
\[
  \begin{array}{rcl}
	\unwind{e_{1}}{J}{c;K} & = & \unwind{e_{1}}{J}{K}	
	\\
	\unwind{e_{1}}{J}{\upd{\beta};K} & = & \unwind{e_{1}}{J}{K}
	\\
	\unwind{e_{1}}{(\stag{z}{h},J)}{\popupd{\beta};K} & = & \unwind{e_{1}}{J}{K}
	\\
	\unwind{e_{1}}{(\stag{e_{1}}{h},J)}{\popupd{\beta};K} & = & (h,\beta,J,K)
  \end{array}
\] 
\end{definition}

\subsect{Implementation of signals}

We compare how our idealized stack machine models features of real signal implementations.

\subsect{Bit vector}
In our machine, $\beta$ stands for the bit vector of installed not currently blocked signals; and $\beta^{0}$ stands for a null bit vector that may be interpreted as "all signals are blocked" or "no signals are installed".
The use of this bit vector almost directly corresponds to the bit maps used in real implementations.
In real implementations, every signal has a default pre-assigned handler. To imitate the same behaviour, in our implementation it is possible to run a command inside of nested blocks in which all signals are bound to their default handlers.

\subsect{Exceptions and signals}
In real implementations (as explained in \cite{de:Dinechin:2000:CEH}, ISO/IEC 14882 \cite{ISO14882:2011,ISO14882:1999}), exception throwing inside of signal handlers is not recommended, due to implementation restrictions.
Moreover, the existing implementation of signals is not block structured.
By contrast, our abstract machine and big-step semantics deal with  block structured signals and allow signal handlers to throw exceptions.

\subsect{Implementation of exception handling}
In real implementations (e.g.: Itanium \cite{IT:CABI:EH}, and as described in \cite{Kerrisk:2010:LinuxProgInter,de:Dinechin:2000:CEH,Bovet:2002:ULK}),  exception handling is implemented by use of stack unwinding. 
Exception handling in our implementation resembles handling in real implementations, except the fact that the abstract machine uses the extra stack $J$ to keep track of block structures, and the $J$ is manipulated by special instructions in the continuation $K$.

\section{Examples of the machine runs}
\label{machinerunsexamples}
We have already seen in previous examples (e.g.: Figure~\ref{perssignhandlbefcom} and Figure~\ref{exnpriorityoversign}) that the big-step semantics gives us block structure for free. This becomes very useful in studying block structured constructs and their interactions.
By contrast, the machine needs to manage block structure explicitly with a help of the stack. 
The examples of corresponding machine runs are given in Figure~\ref{bindintrymachine} and Figure~\ref{tryinbindmachine}.
Please note, the $\popupd{\beta^{0}}^{2}$  stands for $\popupd{\beta^{0}};\popupd{\beta^{0}}$.

The example in Figure~\ref{oneshotsigomegasplit} shows how the big-step syntax makes it easy to address one-shot signals with splitting the bindings. 
On the contrary, the machine needs to perform extra administrative work with the binding tags and the stack to implement one-shot signal handling (Figure~\ref{oneshotsplitmachine}).

One may observe that the abstract machine is more complex than the big-step semantics, as machine needs to deal with many details explicitly.
Overall, we see that the machine is closer to implementations, whereas the big-step semantics is more convenient for abstract reasoning.

\begin{figure}
\begin{eqnarray*}
&&
	\cekhb{\handleexn{(\sigbind{z}{c}{(h \seqcomp\throwexn e)})}{e}{g}}{s_{1}}{\beta^{0}}{\finalcont}{\return}
	\\[\n] 
&\step{}&
	\cekhb{\sigbind{z}{c}{(h \seqcomp\throwexn e)}}{s_{1}}{\beta^{0}}{\stag{e}{g}}{\popupd{\beta^{0}}}
	\\[\n] 
&\step{}&
	\cekhb{c}{s_{1}}{\beta^{0}\sqplus z}{\stag{z}{(h \seqcomp\throwexn e)},\stag{e}{g}}{\popupd{\beta^{0}}^{2}}
	\\[\n] 
&\step{}&
	\cekhb{(h \seqcomp\throwexn e)}{s_{1}}{\beta^{0}}{\stag{z}{(h \seqcomp\throwexn e)},\stag{e}{g}}{\upd{\beta^{0}\sqplus z};c;\popupd{\beta^{0}}^{2}}
	\\[\n] 
&\step{}&
	\cekhb{h}{s_{1}}{\beta^{0}}{\stag{z}{(h \seqcomp\throwexn e)},\stag{e}{g}}{\throwexn e;\upd{\beta^{0}\sqplus z};c;\popupd{\beta^{0}}^{2}}
	\\[\n] 
&\step{}&
	\cekhb{\throwexn e}{s_{2}}{\beta^{0}}{\stag{z}{(h \seqcomp\throwexn e)},\stag{e}{g}}{\upd{\beta^{0}\sqplus z};c;\popupd{\beta^{0}}^{2}}
	\\[\n] 
&\step{}&
	\cekhb{g}{s_{2}}{\beta^{0}}{\finalcont}{\return}
	\\[\n] 
&\step{}&
	\cekhb{\return}{s_{3}}{\beta^{0}}{\finalcont}{\finalcont}
\end{eqnarray*}
\caption{Binding inside of the try block}
\label{bindintrymachine}
\end{figure}

\begin{figure}
\begin{eqnarray*}
&&
	\cekhb{\sigbind{z}{(\handleexn{(\throwexn e)}{e}{g})}{h}}{s_{1}}{\beta^{0}}{\finalcont}{\return}
	\\[\n] 
&\step{}&
	\cekhb{\handleexn{(\throwexn e)}{e}{g}}{s_{1}}{\beta^{0}\sqplus z}{\stag{z}{h}}{\popupd{\beta^{0}}}
	\\[\n]  
&\step{}&
	\cekhb{\throwexn e}{s_{1}}{\beta^{0}\sqplus z}{\stag{e}{g},\stag{z}{h}}{\popupd{\beta^{0}\sqplus z};\popupd{\beta^{0}}} 
	\\[\n] 
&\step{}&
	\cekhb{h}{s_{1}}{\beta^{0}}{\stag{e}{g},\stag{z}{h}}{\upd{\beta^{0}\sqplus z};\throwexn e;\popupd{\beta^{0}\sqplus z};\popupd{\beta^{0}}}
	\\[\n]  
&\step{}&
	\cekhb{\upd{\beta^{0}\sqplus z}}{s_{2}}{\beta^{0}}{\stag{e}{g},\stag{z}{h}}{\throwexn e;\popupd{\beta^{0}\sqplus z};\popupd{\beta^{0}}}
	\\[\n]  
&\step{}&
	\cekhb{\throwexn e}{s_{2}}{\beta^{0}\sqplus z}{\stag{e}{g},\stag{z}{h}}{\popupd{\beta^{0}\sqplus z};\popupd{\beta^{0}}}
	\\[\n]  
&\step{}&
	\cekhb{g}{s_{2}}{\beta^{0}\sqplus z}{\stag{z}{h}}{\popupd{\beta^{0}}}
	\\[\n]  
&\step{}&
	\cekhb{h}{s_{2}}{\beta^{0}}{\stag{z}{h}}{\upd{\beta^{0}\sqplus z};g;\popupd{\beta^{0}}}
	\\[\n]  
&\step{}&
	\cekhb{\upd{\beta^{0}\sqplus z}}{s_{3}}{\beta^{0}}{\stag{z}{h}}{g;\popupd{\beta^{0}}}
	\\[\n]  
&\step{}&
	\cekhb{g}{s_{3}}{\beta^{0}\sqplus z}{\stag{z}{h}}{\popupd{\beta^{0}}}
	\\[\n]  
&\step{}&
	\cekhb{\popupd{\beta^{0}}}{s_{4}}{\beta^{0}\sqplus z}{\stag{z}{h}}{\return}
	\\[\n]  
&\step{}&
	\cekhb{\return}{s_{4}}{\beta^{0}}{\finalcont}{\finalcont}
\end{eqnarray*}

\caption{Exception handling inside of the binding}
\label{tryinbindmachine}
\end{figure}

\begin{figure}
\begin{eqnarray*}
&&
	\cekhb{\sigonce{z}{(c_{1} \seqcomp c_{2})}{c_{h}}}{s_{1}}{\beta^{0}}{\finalcont}{\return}
	\\[\n] 
&\step{}&
	\cekhb{c_{1} \seqcomp c_{2}}{s_{1}}{\beta^{0}\sqplus z}{\ttag{z}{h_{1}}{0}}{\popupd{\beta^{0}};\return}
	\\[\n] 
&\step{}&
	\cekhb{c_{1}}{s_{1}}{\beta^{0}\sqplus z}{\ttag{z}{h_{1}}{0}}{c_{2};\popupd{\beta^{0}};\return}
	\\[\n] 
&\step{}&
	\cekhb{h_{1}}{s_{1}}{\beta^{0}}{\ttag{z}{h_{1}}{1}}{\upd{\beta^{0}};c_{1};c_{2};\popupd{\beta^{0}};\return}
	\\[\n] 
&\step{}&
	\cekhb{\upd{\beta^{0}}}{s_{2}}{\beta^{0}}{\ttag{z}{h_{1}}{1}}{c_{1};c_{2};\popupd{\beta^{0}};\return}
	\\[\n] 
&\step{}&
	\cekhb{c_{1}}{s_{2}}{\beta^{0}}{\ttag{z}{h_{1}}{1}}{c_{2};\popupd{\beta^{0}};\return}
	\\[\n] 
&\step{}&
	\cekhb{c_{2}}{s_{3}}{\beta^{0}}{\ttag{z}{h_{1}}{1}}{\popupd{\beta^{0}};\return}
	\\[\n] 
&\step{}&
	\cekhb{\popupd{\beta^{0}}}{s_{4}}{\beta^{0}}{\ttag{z}{h_{1}}{1}}{\return}
	\\[\n] 
&\step{}&
	\cekhb{\return}{s_{4}}{\beta^{0}}{\finalcont}{\finalcont}
\end{eqnarray*}

\caption{Signal binding and seq. composed commands}
\label{oneshotsplitmachine}
\end{figure}

\section{Conclusions}
\label{conclusions}

The present paper idealizes signal handling in combination with the more familiar exception handling to focus on some of their semantic and logical features.
The semantics of one-shot handlers is reminiscent of linearly-used continuations~\cite{LinUCHOSC} and the resource usage in separation logic~\cite{reynoldslicssep}. The way we have treated signal bindings in the big-step semantics borrows ideas from linear logic. Recall that we write
\[
\bs SO{s_{1}}{c}{s_{2}}
\] 
for a judgement involving a persistent signal binding $S$ and a one-shot signal binding $O$. As we have illustrated with the examples in Section~\ref{secexamples}, the signal binding $S$ can be shared between two commands $c_1$ and $c_2$ in a sequential composition, whereas $O$ has to be split into disjoint parts $O_1$ and $O_2$. This splitting prevents a one-shot signal handler from being re-used and makes it a linear resource just like the contexts in a linear logic. In fact, Dual Intuitionistic Linear Logic~\cite{dill}  has two zones $\Gamma$ and $\Delta$ in the context, one which allows sharing and one which does not, as in the following rule that shares $\Delta$ and splits $\Gamma$:
\[
\infer{\Gamma_1;\Delta \vdash M:A\multimap B \qquad \Gamma_2;\Delta\vdash N:A}{\Gamma_1,\Gamma_2;\Delta \vdash M\,N:B}
\]
We are not aware of previous operational semantics for signals, although
Feng, Shao, Guo and Dong
\cite{certifyinginterrupts} presents a program logic for assembly language with interrupts, which are analogous to signals at the hardware level.

Hutton and Wright \cite{HuttonW07} study interruptions as asynchronous exceptions.
By contrast, signals are a software alternative to  hardware interrupts, where signal handlers could be addressed as asynchronous subroutine calls. 

Signals have been part of the long evolution of Unix, and are correspondingly complex. To implement block-structured signal handling and integrate it with exceptions, the present signal mechanism may have to be revisited.
The present implementations pose severe restrictions on programmers, for instance on using non-local control in a handler. Removing such implementation restrictions would enable natural programming idioms. In further work, we hope to build on the operational semantics presented here for proving soundness of a Hoare logic for signals.

The formal connection between the big-step operational semantics and the signals abstract machine remains to be established. We conjecture that they are observationally equivalent and that this may be proved by way of a simulation
relation.

\bibliographystyle{eptcs}

\end{document}